\documentclass[aps,twocolumn,pra,groupedaddress,superscriptaddress,amsmath,amssymb,showpacs,noeprint]{revtex4-2}

\usepackage{float}
\usepackage[table,dvipsnames]{xcolor}
\usepackage{dcolumn}
\usepackage{bm}
\usepackage{graphicx}
\usepackage{amsmath}
\usepackage{amsfonts}
\usepackage{amssymb}
\usepackage{amsthm}
\usepackage{amstext}
\usepackage{amsbsy}
\usepackage{amsopn}
\usepackage{amscd}
\usepackage{amsxtra}
\usepackage{soul}

\usepackage{hyperref}
\usepackage{booktabs}
\usepackage{multirow}

\newcolumntype{C}{>{$}c<{$}}
\AtBeginDocument{
\heavyrulewidth=.08em
\lightrulewidth=.05em
\cmidrulewidth=.03em
\belowrulesep=.65ex
\belowbottomsep=0pt
\aboverulesep=.4ex
\abovetopsep=0pt
\cmidrulesep=\doublerulesep
\cmidrulekern=.5em
\defaultaddspace=.5em}
\newcommand{\Hilbert}{\mathcal{H}}
\newcommand{\Mc}[1]{\mathcal{#1}}

\newcommand{\setR}{\mathbb{R}}

\newcommand{\setC}{\mathbb{C}}

\newcommand{\bra}[1]{\langle #1 |}
\newcommand{\ket}[1]{|#1\rangle }

\newcommand{\ii}{i}
\newcommand{\A}{\hat a}
\newcommand{\ad}{\hat a^\dagger}

\newcommand{\sigp}{\hat{\sigma}_+}
\newcommand{\sigm}{\hat{\sigma}_-}
\newcommand{\sigz}{\hat{\sigma}_z}

\newcommand{\coe}{\alpha}
\newcommand{\squ}{\beta}

\everymath{\displaystyle}

\begin{document}

\title{Exploring the limits of the generation of non-classical states of spins coupled to a cavity by optimal control}

\author{Quentin Ansel}
\email{quentin.ansel@univ-fcomte.fr}
\affiliation{Institut UTINAM - UMR 6213, CNRS, Universit\'{e} Bourgogne Franche-Comt\'{e}, Observatoire des Sciences de l'Univers THETA, 41 bis avenue de l'Observatoire, F-25010 Besan\c{c}on, France}

\author{Dominique Sugny}
\affiliation{Laboratoire Interdisciplinaire Carnot de Bourgogne, CNRS UMR 6303, Universit\'{e} Bourgogne Franche-Comt\'{e}, BP 47870, F-21078 Dijon, France}

\author{Bruno Bellomo}
\affiliation{Institut UTINAM - UMR 6213, CNRS, Universit\'{e} Bourgogne Franche-Comt\'{e}, Observatoire des Sciences de l'Univers THETA, 41 bis avenue de l'Observatoire, F-25010 Besan\c{c}on, France}

\begin{abstract}

We investigate the generation of non-classical states of spins coupled to a common cavity  by means of a
collective driving of the spins. We propose a control strategy using specifically designed series of short coherent and squeezing pulses, which have the key advantage of being experimentally implementable  with the state-of-the art techniques. The parameters of the control sequence are found by means of optimization algorithms. We consider the cases of two and four spins, the goal being either to reach a well-defined target state or a state maximizing a measure of non-classicality. We discuss the influence of cavity damping and spin offset on the generation of non-classical states. We also explore to which extent squeezing fields help enhancing the efficiency of the control process.
\end{abstract}

\maketitle

\section{Introduction}

Physical systems exhibiting non-classical features such as quantum coherence~\cite{coherence2017} and quantum entanglement~\cite{RMPentanglement,entanglement2009} play a key role in the development of quantum technologies~\cite{QT}. These features are precious resources that must be engineered and controlled in the most efficient way~\cite{glaser_training_2015} because of their fragility due to decoherence processes induced by the interaction with the environment~\cite{BookBreuer2007,macroquantum2018}. In this framework, the generation of entangled states has been widely studied theoretically and experimentally in different contexts~\cite{stockton_deterministic_2004,taylor_solid-state_2005,friesen_efficient_2007,neumann_multipartite_2008,simmons_entanglement_2011,mcconnell_generating_2013, Bellomo2015, estarellas_robust_2017, Bellomo2019,Bellomo2020}. In view of concrete and experimental implementations, it can be crucial to know how and to which extent non-classical states can be generated in a specific setup.

This question can be addressed by quantum Optimal Control Theory (OCT)~\cite{brysonbook,alessandrobook,glaser_training_2015,bonnard_optimal_2012,PRXQuantumsugny}, which has become nowadays  one of the main tools to design external electromagnetic fields able to control quantum dynamics. OCT was first applied in molecular systems~\cite{brif:2010,RMP:rotation} and Nuclear Magnetic Resonance (NMR)~\cite{kobzar:2008,khanejaspin,lapertprl,skinner:2005} to control chemical reactions or to prepare spin states for spectroscopy and imaging purposes~\cite{conolly:1986,vinding:2012,maximov,bonnard,lapert_exploring_2012}. OCT is by now a central pillar in quantum technologies~\cite{QT,glaser_training_2015,koch_controlling_2016}. The generation of entangled spin states using optimal magnetic fields has been studied extensively in NMR and trapped ions~\cite{platzer_optimal_2010,nebendahl_optimal_2009}, but less attention has been paid to Cavity-QED (CQED) experiments, where a large number of spins interacts through a quantized cavity mode, giving rise to collective quantum effects, such as superradiance and subradiance~\cite{andreev_collective_1980,gross_superradiance:_1982,gegg_superradiant_2018}. The generation of spin entanglement in CQED has been demonstrated experimentally using feedback control~\cite{brakhane_bayesian_2012}. However, considerable progress in open-loop control processes are  still needed to exploit the full potential of CQED experiments. Finally, we note that many physical and technological applications can be derived from the control of non-classical spin states. Among others, we mention the study of collective behaviors~\cite{filipp_preparation_2011} and quantum information processing~\cite{wendin_quantum_2017}.

Recently, an increasing interest emerged in high sensitivity Electron Spin Resonance (ESR) experiments~\cite{bienfait_magnetic_2016,probst_inductive-detection_2017,morton_storing_2018,haikka_proposal_2017}. Differently from trapped-ion experiments, atoms cannot be addressed individually, and only collective transformations can be achieved. This key constraint reduces considerably the way in which entangled states can be generated. In recent ESR experiments, the cavity field can be modified using both coherent and squeezed states of light~\cite{bertetPRX17}. In this context, note that squeezed states have been considered only in a few theoretical~\cite{gardiner_inhibition_1986,gardiner_driving_1994,shahmoon_qubit_2009,ficek_entangled_2002,arenz20,nori2015} and experimental~\cite{Burd1163} quantum studies.

In Ref.~\cite{ansel_optimal_2018-1}, some of us considered the optimal control of a spin ensemble coupled to a single lossy cavity mode. The study was restricted to the bad-cavity regime, and only coherent control fields were used. In this regime, a semi-classical approximation can be used to neglect entanglement. Many papers studied this kind of systems both on the theoretical and experimental sides (see \cite{Krimer_2019_1,Krimer_2019_2} and references therein). 
	
In this paper, we go beyond the scope of the previous study \cite{ansel_optimal_2018-1} by exploring the generation of non-classical spin states that violate the semi-classical approximation, in a setup similar to those encountered in high sensitivity ESR. We consider an ensemble of spins coupled to a single lossy cavity mode and we study how to drive the spins toward a specific state, typically in a minimum time, using both coherent and squeezing controls acting on the cavity. We propose a control strategy using specifically designed series of short pulses which could be experimentally implemented with the state-of-the-art techniques. We consider systems with two and four spins for which the control objective is to generate, respectively, symmetric and antisymmetric states, or to maximize a measure of non-classicality. We explore to which extent optimal control allows one to tackle these objectives keeping into account also the role of spin offset inhomogeneities and cavity damping. In addition, we show in which conditions squeezing fields can help enhancing the efficiency of control processes. Finally, we investigate the robustness of some optimized pulse sequences against different experimental uncertainties on the coupling strength to the cavity, the spin offsets, and the cavity damping. Note that contrary to feedback control protocols, which are based on the updated knowledge of the system state to adjust the control strategy, we do not have direct access to the spin state. Without this key resource, the generation of entanglement in an open-loop configuration is even more challenging \cite{glaser_training_2015}.

The paper is organized as follows. In Sec.~\ref{sec:model}, we present the model under study and the figures of merit we will use to go beyond the scope defined by the semi-classical approximation. In Sec.~\ref{Sec:Preliminary considerations on the control of our system} we present the pulse sequence parametrization, the numerical optimization procedure, and we provide a preliminary analysis of the system controllability. In Sec.~\ref{sec:2_spins}, we investigate the generation of symmetric and antisymmetric states for an ensemble of two spins, whereas in Sec.~\ref{sec:many_spins} we extend the analysis to four spins also using a measure of non-classicality. Conclusion and prospective views are given in Sec.~\ref{sec:conclusion}. Some technical details, a few optimized pulse sequence parameters and additional numerical results are reported in the appendices.

\section{Model and figures of merit to go beyond the semi-classical approximation}\label{sec:model}

\subsection{Description of the model}
We consider an ensemble of $N_s$ spin-1/2 particles coupled via a Jaynes-Cummings interaction to a single cavity mode which is driven by some external fields.
The cavity is assumed to have losses due to its coupling to a zero-temperature environment. Under the rotating wave~\cite{haroche_exploring_2006} and the fixed dissipator approximations~\cite{giorgi_microscopic_2020} the master equation in the rotating frame at the cavity frequency $\omega_c$ is given by
\begin{equation}
\frac{d \hat \rho}{dt} = -\frac{\ii}{\hbar} [\hat H(t),\hat \rho] + \kappa\left[ \A \hat \rho \ad -\frac{1}{2}\left(\ad\A \hat \rho + \hat \rho \ad\A\right) \right],
\label{eq:master_eq}
\end{equation}
where $\kappa$ is the damping rate of the cavity, and $\A,~\ad$ are respectively the annihilation and creation operators. The Hamiltonian $\hat H(t)$ can be expressed as
\begin{equation}
\begin{split}
\hat H(t) =& \underbrace{
\ii \hbar \left[ \coe(t)\ad -\coe(t)^*\A \right] + \squ (t) \frac{\ii\hbar}{2}\left[(\ad)^2 -(\A)^2 \right] }_{\hat H_C} \\
& + \underbrace{ \sum_{n=1}^{N_s} \frac{\hbar \Delta_n}{2}\sigz^{(n)} + \hbar g\sum_{n=1}^{N_s} \left[\ad \sigm^{(n)}+\A \sigp^{(n)}\right]}_{\hat H_0},
\end{split}
\label{eq:full_Hamiltonian}
\end{equation}
where $\sigp^{(n)},~\sigm^{(n)},~\sigz^{(n)}$, and $\Delta_n\equiv\omega_s^{(n)}-\omega_c$ are, respectively, the Pauli operators associated with the $n$-th spin and its detuning (also called offset) with respect to the cavity frequency, being $\omega_s^{(n)}$ the frequency of the $n$-th spin. The parameter $g$ is the coupling strength between the spins and the cavity mode, which is assumed to be real, positive, and equal for all the spins.  The control fields depending on $\coe(t) \in \setC$ and $\squ(t) \in \setR$ (see Appendix~\ref{sec:appendix The short pulse limit and the squeezing control} for a justification of the latter choice) generate respectively coherent and squeezed states when they are applied to the cavity vacuum. For these reasons, they are referred as coherent and squeezing controls. Notice that we assume that $|\coe|$, $|\squ|$, $g$, and $|\Delta_n|$ are much smaller than the cavity frequency $\omega_c$ in order to use the rotating wave~\cite{haroche_exploring_2006} and the fixed dissipator approximations~\cite{giorgi_microscopic_2020}. The latter allows us to use the same dissipator describing the cavity losses in the absence of spins. In the following, we adopt the notation $\hat J_a = \sum_{n=1}^{N_s} \hat \sigma_a^{(n)}$, with $a=+,-,z$.

The Hilbert space of the total system is given by $\Hilbert = \Mc F \otimes \setC ^ {2N_s}$, where $\Mc F$ is the Fock space of the cavity mode, which is truncated in numerical simulations. The pure states can be expressed as $\ket{n,\psi_S}$ where $n$ is the number of excitations in the cavity and $|\psi_S\rangle$ is a pure state of the spin ensemble. For a two-spin system, we use the basis \{$\ket{G}=\ket{\downarrow,\downarrow}$, $\ket{A}= (\ket{\uparrow,\downarrow}-\ket{\downarrow,\uparrow})/\sqrt{2}$, $\ket{S}= (\ket{\uparrow,\downarrow}+\ket{\downarrow,\uparrow})/\sqrt{2}$,  $\ket{E}=\ket{\uparrow,\uparrow}$\}, where $\ket{\downarrow} $ and $\ket{\uparrow}$ are, respectively, the ground and the excited levels of the $\sigz^{(n)}$ operator. Similar states can be defined for an even number of more than two spins~\cite{arecchi_atomic_1972}. In particular, for a four-spin system, we still use $\ket{A}$ and $\ket{S}$ to denote, respectively, the antisymmetric and symmetric states with two excitations.

\subsection{Dynamics beyond the semi-classical approximation}

Many studies in the literature concerned the behavior of an ensemble of atoms in a lossy cavity in the limit when a semi-classical (or mean field) approximation can be performed to study the system evolution~\cite{ansel_optimal_2018-1,Krimer_2019_1,Krimer_2019_2}. This approximation is connected to the notion of cumulant. In the case of two arbitrary operators $\hat A$ and $\hat B$, it is defined by $\langle \hat A \hat B \rangle_c \equiv \langle \hat A \hat B \rangle - \langle \hat A \rangle \langle\hat B \rangle$~\cite{kubo_generalized_1962}. If the operators act on two different subsystems of a larger system and if the density matrix of these two subsystems is a separable state with classical-classical correlations \cite{adesso_2015} of the form $\hat \rho=\sum_{i,j} p_{i}^{(1)}p_{j}^{(2)} \ket{i}\bra{i} \otimes \ket{j}\bra{j}$, then, the cumulant is equal to zero. Moreover, any entangled state is characterized by a non-zero cumulant. Note that $ p^{(1)}$ and $p^{(2)}$ are probability distributions and $\{\ket{i}\}$, $\{\ket{j}\}$ are orthonormal bases for the two subsystems.

In the semi-classical limit, the dynamics of the many-body system can be expressed in terms of a few observable mean values. This approximation is valid when some cumulants are negligible and set to zero. For example, the equation of motion for $\langle \hat \sigma^{(n)}_z \rangle$ is given by $d_t \langle \hat \sigma^{(n)}_z \rangle = 2 \ii g \left( \langle \hat \sigma_-^{(n)} \hat a^\dagger \rangle - \langle \hat \sigma_+^{(n)} \hat a\rangle\right)$, which is approximated into  $d_t \langle \hat \sigma^{(n)}_z \rangle = 2 \ii g \left( \langle \hat \sigma_-^{(n)} \rangle \langle\hat a^\dagger \rangle - \langle \hat \sigma_+^{(n)}\rangle \langle \hat a\rangle\right)$. We refer to Refs.~\cite{ansel_optimal_2018-1,Krimer_2019_1,Krimer_2019_2} for the presentation of the semi-classical approximation in the case of a spin ensemble coupled to a cavity.

A main aspect of this paper is to go beyond the validity regimes of the semi-classical approximation, by generating non-classical states for which this approximation does not hold. On the basis of the above considerations, maximizing a figure of merit defined by the cumulants of spin operators is expected to allow us to generate spin states that violate the semi-classical approximation. In our numerical optimizations we will make use of two tools to obtain such kind of states.

\noindent\textbf{Fidelity to the target state.} The first one is the fidelity of the generated state at the control time $t_{f}$, $\hat \rho(t_f)$, with respect to some target state $\hat \rho_{\mathrm{ts}}$, which will be chosen as an highly entangled state. This fidelity at an arbitrary time $t$ is defined as \cite{yuan_fidelity_2017}:
\begin{equation} \label{eq:fidelity}
F(t)=\left(\text{Tr}\left[\sqrt{\sqrt{\hat \rho_\mathrm{ts}} \hat \rho (t) \sqrt{\hat \rho_\mathrm{ts}}}\right]\right)^2.
\end{equation}
In the following, we denote by $F$ the value of this quantity at time $t_f$.
Maximizing the fidelity with respect to a specific state is expected to be difficult for $N_s >2$. Indeed, when the number of spin increases there are many different entangled states in the Hilbert space, so that it could be advantageous to introduce a figure of merit that does not depend explicitly on a target state.

\noindent\textbf{Measure of non-classicality.} This brings us to the second tool, which is a measure of non-classicality based on the cumulant associated to all the possible products of the form $\hat \sigma_+^{(n)} \hat \sigma_-^{(m)}$. A quantity of this kind has the advantage of being computable with a single time integration, and the definition does not depend on the system dimension.
In particular, in order to have a figure of merit bounded by 1, we define at an arbitrary time $t$,
\begin{align}
\label{eq:def_cumulant_FoM}
&\Mc C (t) = \frac{8}{N_s^2}\sum_{n=1}^{N_s}\sum_{m=1}^{n}\langle \sigp^{(n)} \sigm^{(m)}\rangle_c (t) \\
&= \frac{8}{N_s^2} \sum_{n=1}^{N_s} \nonumber \sum_{m=1}^{n} \left(\langle\sigp^{(n)} \sigm^{(m)}\rangle (t) - \langle \sigp^{(n)}\rangle (t) \langle \sigm^{(m)}\rangle (t)\right).
\end{align}
In the following, we denote by $\Mc C$ the value of this quantity at time $t_f$.
A spin-state verifying exactly the semi-classical approximation is such that $\langle \hat \sigma_+^{(n)} \hat \sigma_-^{(m)} \rangle_c = 0, ~ \forall n,m$, and thus $\Mc C=0$ for such a state. Moreover, $\Mc C =1$ for a $\ket{N_s/2,0}$ Dicke state (we have used the notation $\ket{J, M}$, where $J$ is the total angular momentum and $M$ its projection on the $z$ axis)~\cite{gross_superradiance:_1982}. Indeed, we have  $\bra{N_s/2,M}  {\hat \sigma}^{(n)}_\pm \ket{N_s/2,M}=0$ and $\bra{N_s/2,M}\sum_{n=1}^{N_s}\sum_{m=1}^{n} \sigp^{(n)} \sigm^{(m)} \ket{N_s/2,M}=[(N_s/2)^2-M^2]/2$. This latter is equal to $N_s^2/8$ for the symmetric state, which is characterized by $M=0$.
On the other hand, $\Mc C =-0.33$ for a totally antisymmetric state of four spins. By construction, we expect that  in an optimization procedure, maximizing $\Mc C$ and $-\Mc C$ would lead, respectively, to superradiant and subradiant states. Maximizing a specific measure, such as the one defined in Eq.~\eqref{eq:def_cumulant_FoM}, will allow us to increase, in general, the entanglement without an explicit choice of target state. For a given system, the value of $\Mc C$ for which the semi-classical approximation does not hold can be estimated. This can be achieved by comparing the dynamics of mean value of operators, which are both computed using the semi-classical and the full quantum models. As example, we choose the relaxation of an excited spin ensemble toward their ground state. For the four-spin system studied in Sec.~\ref{sec:many_spins}, a good agreement between the quantum and the classical models is observed for $\Mc C \lesssim 0.15$.

\noindent\textbf{Entanglement quantifier.} Another possible approach is to use an entanglement quantifier as a figure of merit to maximize~\cite{plenio_introduction_2007}.
However, entanglement measures are generally computationally expensive and difficult to extend to many-body systems. In the context of a spin ensemble coupled to a single cavity mode, it is more convenient to use an indicator of entanglement based on operator mean values. They may not verify all the required properties of an entanglement measure, but they generally have a simple physical interpretation, and they provide quantitative information on the amount of non-classicality of a state. For example, in addition of the specific measure defined in Eq.~\eqref{eq:def_cumulant_FoM}, we can mention the cooperativity fraction~\cite{mascarenhas_cooperativity_2013}, the collectivity measure~\cite{gegg_superradiant_2018}, the lower bound of multipartite concurrence~\cite{platzer_optimal_2010}, or the correlator $\langle \hat J_+ \hat J_- \rangle_{\mathrm{corr}}$ (see Appendix~\ref{sec:state_properties_relaxation} for its definition)~\cite{temnov_superradiance_2005,ficek_entangled_2002,andreev_collective_1980,gross_superradiance:_1982}.

\section{Preliminary considerations on the control process}\label{Sec:Preliminary considerations on the control of our system}

In this section, we deal with several preliminary aspects concerning the control of our system before presenting the numerical results.

\subsection{Approximation and control mechanisms}

The control part of the unitary term of the dynamic of Eq.~\eqref{eq:master_eq} is modeled by the Hamiltonian $\hat H_C$ which is defined in Eq.~\eqref{eq:full_Hamiltonian} and which must respect some  specific constraints. Indeed, if the control fields have a large amplitude for a long control time, the fixed dissipator approximation might not be valid anymore~\cite{giorgi_microscopic_2020}. Additionally, arbitrary time-dependent controls may be difficult to implement experimentally and they could lead to experimental or numerical artifacts. In order to avoid such problems, we introduce the short pulse limit, which is realized  approximatively using bump pulses~\cite{ansel_optimal_2018,ansel_optimal_2018-1,probst_shaped_2019}. We refer to Appendix~\ref{sec:short_pulse_limit} for technical details about this approximation. This subsection is limited to a summary of its content.

Bump pulses are parameterized by a smooth analytic function. They allow us to replace the coherent control [terms with $\coe$ in Eq.~\eqref{eq:full_Hamiltonian}] by:
\begin{equation}
\frac{\hbar}{2}\left[\theta_{k,x} \delta(t-\tau_{k,x}) \hat J_x + \theta_{k,y} \delta(t-\tau_{k,y}) \hat J_y\right],
\end{equation}
where $\theta_{k,\mu}$ is the flip-angle number $k$ along the axis $\mu$, and $\tau_{k,\mu}$ are times where pulses are applied. This approximation is justified when the pulse duration is negligible in front of $1/g$. Moreover, the pulse is constructed in order to rotate all the spins of an angle $\theta_{k,\mu}$. The same approximation cannot be performed with the squeezing field (due to quadratic terms in cavity-field operators). Furthermore, coherent and squeezing controls do not commute~\cite{nieto_holstein-primakoff/bogoliubov_1997}. Therefore, the short pulse approximation is valid only if the two control parameters are not simultaneously switched on. Finally, we obtain the following approximated control Hamiltonian:
\begin{equation}
\hat H_C(t)\approx u_x(t)\underbrace{\frac{\hbar}{2} \hat J_x }_{\hat V_x}+u_y(t) \underbrace{\frac{\hbar}{2} \hat J_y}_{\hat V_y}+ \squ (t)\underbrace{ \frac{\ii\hbar}{2}\left[(\ad)^2 -(\A)^2 \right] }_{\hat V_s},
\end{equation}
where $u_x = \sum_k \theta_{k,x} \delta(t-\tau_{k,x})$ and $u_y = \sum_k \theta_{k,y} \delta(t-\tau_{k,y})$ are sequences of $\delta$-pulses, and we have denoted by $\hat V_x$, $\hat V_y$, and $\hat V_s$ the interaction operators associated respectively with the controls $u_x$, $u_y$, and $\beta$.

Differently from the case of coherent control, we cannot associate the squeezing control to an angle of rotation. However, there is an interesting interpretation in terms of effective coupling strength with the cavity. We refer to Appendix~\ref{sec:squeezing control} for further details.

\subsection{Parametrization of the pulse sequence and numerical optimization}

We consider a specific parametrization of the control law displayed in Fig.~\ref{fig:fig_pulses_packages}, which is made up of a series of pulses called \emph{a pulse package}.

Note that the cumulative duration of a \emph{pulse package} should be very small compared to $1/g$, in order to satisfy the short pulse limit. The pulse sequence can be formally written using the evolution operator of the Schr\"odinger equation. By using the fact that a Dirac distribution allows one to transform a time ordered exponential into an ordered product of exponential, we have:
\begin{figure}
\includegraphics[width=\columnwidth ]{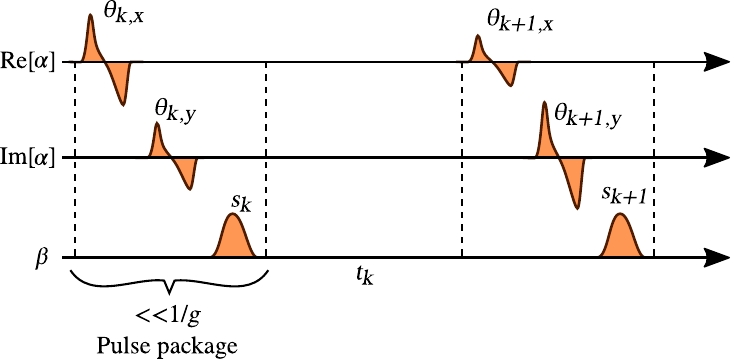}
\caption{Scheme of the pulse sequence which corresponds to a series of pulse package made up of two coherent bump pulses and a squeezed control.}
\label{fig:fig_pulses_packages}
\end{figure}
\begin{equation}
\hat U = \mathbb{T} \prod_{k=1}^{M}e^{-\frac{\ii}{\hbar} t_k \hat H_0} \underbrace{e^{-\frac{\ii}{\hbar}  s_k \hat V_s}e^{-\frac{\ii}{\hbar}  \theta_{k,y} \hat V_y}e^{-\frac{\ii}{\hbar}  \theta_{k,x} \hat V_x}}_{\text{a pulse package}},
\label{eq:evolution_operator_unitary}
\end{equation}
where $\mathbb{T}$ is the time ordering operator and the control law is given by a set of values $\{\theta_{k,x},\theta_{k,y},s_k,t_k\}_{k=1\cdots M}$, being $M$ the number of pulse packages. The variable $s_k$ is the integral over the time of $\beta(t)$ during the duration of the $k$-th pulse [this corresponds to the area under the $\beta$ curve in Fig.~\ref{fig:fig_pulses_packages} during one pulse]. The optimizations are performed on a bounded set in which $\theta_{k,i} \in [0, 2\pi]$ and $s_k\in [-s_{\mathrm{max}},s_{\mathrm{max}}]$. The bound $s_{\mathrm{max}}$ is chosen so that it does not introduce numerical artifacts on the truncated Fock-space. In the numerical computations, we use $s_{\mathrm{max}} \in [0.5,2]$. In Appendix~\ref{sec:sequence_parameters}, we give the range of the system dimensions considered in our simulations and we provide the parameters of some pulse sequences used in this study. For sake of completeness, we also investigate the robustness of some optimized pulse sequences against uncertainties on the coupling strength, spin offsets, and cavity damping.
Equation~\eqref{eq:evolution_operator_unitary} is adapted to the dissipative case by replacing $-\ii \hat H_0 /\hbar$ by the Lindblad super-operator associated to the dissipative dynamics in the absence of controls, and by changing the interaction operators $\hat V_i$ by commutators of the form $[\hat V_i, \cdot]$. The Lindblad equation is integrated with a standard split operator algorithm. In this paper, we limit the number of pulse packages to $M=7$, which leads to a maximum number of 28 parameters to optimize. However, numerical simulations show that $M=5$ is generally enough to reach the target state with a very high fidelity.

Some details about the employed numerical optimization procedure, which is based on a two-step protocol, are reported in Appendix~\ref{sec:sequence_parameters}. This procedure involves a preliminary application of the \emph{NMaximize} function of \textit{Mathematica}, followed by the employment of a home-made version of the JAYA algorithm~\cite{venkata_rao_jaya:_2016}.

\subsection{System controllability}
\label{sec:controlability}

The controllability of the Jaynes-Cummings model (or of the generalized Tavis-Cummings model) has been studied extensively during the past years~\cite{keyl_controlling_2014,pinna_controllability_2018,mirrahimi_controllability_2005,hofmann_controlling_2017}. It has been emphasized that the infinite-dimensional system is approximatively controllable in a generic situation.
In this subsection we study if the coherent and squeezing fields are sufficient to control our system. To simplify this analysis, we neglect the relaxation and we consider a finite-dimensional approximation of the total quantum system by truncating the Fock space of the cavity mode.

In particular, we show the exact controllability of a truncated system, following the methods presented in Refs.~\cite{schirmer_complete_2001,altafini_controllability_2002}.
For that purpose, we verify that the group $SU(N)$ (with $N$ the dimension of the truncated space) of evolution operators can be generated by the set of operators $(\hat H_0, \hat V_x)$, $(\hat H_0, \hat V_x, \hat V_y)$, or $(\hat H_0, \hat V_x, \hat V_y, \hat V_s)$. This is performed by calculating the dimension of the Lie algebra generated by the recursive commutators of these operators. If the dimension of the Lie algebra converges to dim$(\mathfrak{su}(N))$, being $\mathfrak{su}(N)$ the Lie algebra of  $SU(N)$, then the system is controllable. We note that the case $(\hat H_0, \hat V_y)$ is equivalent to the case $(\hat H_0, \hat V_x)$ for symmetry reasons.

\begin{table}[t]
\begin{center}
\begin{tabular}{c|ccccccc} \toprule
Commutator order   &  0 & 1 & 2 & 3 & 4 & 5 & 6 \\ \midrule
$(\hat H_0, \hat V_x)$           & 2 & 3 & 5 & 10 & 34 & 153 & 153  \\
$(\hat H_0, \hat V_x, \hat V_y)$           & 3 & 6 & 12 & \cellcolor{blue!25} 44 & 288 & \cellcolor{gray!20} 323 &  \cellcolor{gray!20} 323 \\
$(\hat H_0, \hat V_x, \hat V_y, \hat V_s)$ & 4 & 8 & 21 & \cellcolor{blue!25}138 & \cellcolor{gray!20}323 & \cellcolor{gray!20} 323 & \cellcolor{gray!20} 323\\ \bottomrule
\end{tabular}
\end{center}
\caption{Dimensions of the Lie algebra spanned by recursive commutators of the Hamiltonians. The commutator order is defined as the maximum number of commutators taken into account in the computation. The order 0 corresponds to the terms $\hat V_i$, being $i\in \{0,x,y,s\}$ and $\hat V_0 \equiv \hat H_0$, the order 1 to the terms $\{ \hat V_i, [\hat V_i, \hat V_j]\}$, the order 2 to the terms $\{\hat V_i, [\hat V_i, \hat V_j], [ [\hat V_i, \hat V_j],\hat V_k]\}$, etc. In this example, the Fock space is truncated to five excitations, and the dimension of the spin Hilbert space is three (two spins at resonance represented in the Dicke basis by removing the antisymmetric state since at resonance it is dynamically decoupled to other states). We have dim$\Hilbert = 18$ and dim$(\mathfrak{su}(18))=N^2-1 = 323$.
}
\label{tab:dim_lie_algebra}
\end{table}

An example of calculation is presented in Tab.$~$\ref{tab:dim_lie_algebra} for two spins at resonance ($\Delta_1=\Delta_2=0$) and the Fock space truncated to five excitations.  Similar results are obtained both for smaller systems and for configurations out of resonance.

Two observations can be made. The first one is that when $\Delta_1=\Delta_2=0$, the system is controllable only in the presence of the two coherent controls along the $x$ and $y$ directions (it is not controllable with a single coherent control). The second point concerns a noticeable improvement in the generation of the dynamical algebra with a squeezing control. Convergence is obtained with a smaller order of commutators (improvement of a factor 3.1 at order 3 where the dimension of the algebra is 138 with the squeezing, instead of 44 without squeezing). The commutators can be introduced explicitly in the expression of the evolution operator using the Baker-Campbell-Hausdorff formula, where they appear as perturbation terms.  The elements of the algebra generated by high order commutators require, in general, large control times in order to have a non-negligible impact on the dynamics. Then, decreasing the commutator order may lead to a reduced control time.

However, this qualitative interpretation is limited by two key aspects. The original system is infinite dimensional and some target states are generated only approximatively in a finite time. In addition, we have no information about the states that can be reached more easily with a squeezing control. In other words, the control  strongly depends on the states to attain, and a case-by-case study is therefore necessary to solve these different points.

\section{Generation of two-spin symmetric / antisymmetric States}
\label{sec:2_spins}

In this section, we consider a two-spin system and we focus on the generation of symmetric and antisymmetric states, using coherent and squeezing controls. The additional constraint that the cavity goes back to its ground state at the final time is accounted for. This constraint is experimentally interesting to detect spin response when the control is switched off.

\subsection{Optimal control without relaxation}
\begin{figure}[t]
\includegraphics[width=\columnwidth]{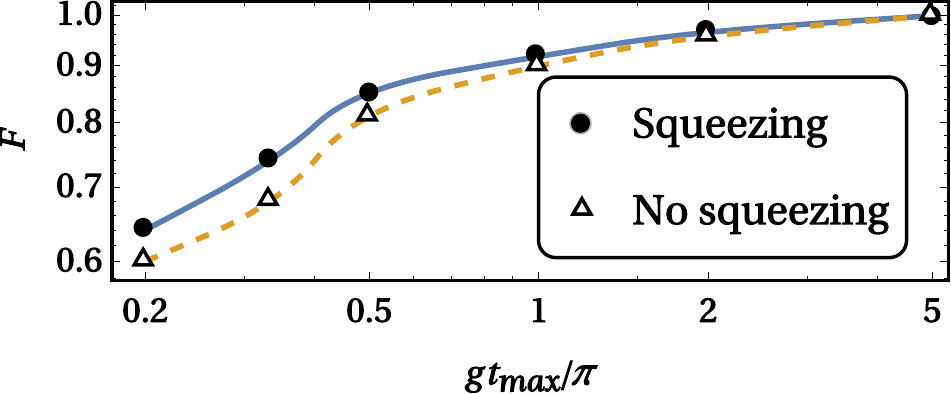}
\caption{Two-spin system without damping ($\kappa=0$) at resonance. Maximum fidelity with respect to the symmetric state $\ket{S}$ as a function of $gt_{\mathrm{max}}$. The black dots and the solid blue curve refer to the controls with squeezing, while the open triangles and the orange dashed curve correspond to the ones without squeezing. Interpolated curves are used to guide the reading. The pulse sequence used to generate the point with $gt_{\text{max}}/\pi =5$ (no squeezing) is given in Tab.~\ref{tab:sequence_parameters}.}
\label{fig:squeezing_speed_up}
\end{figure}

\begin{figure}[h]
\includegraphics[width=\columnwidth]{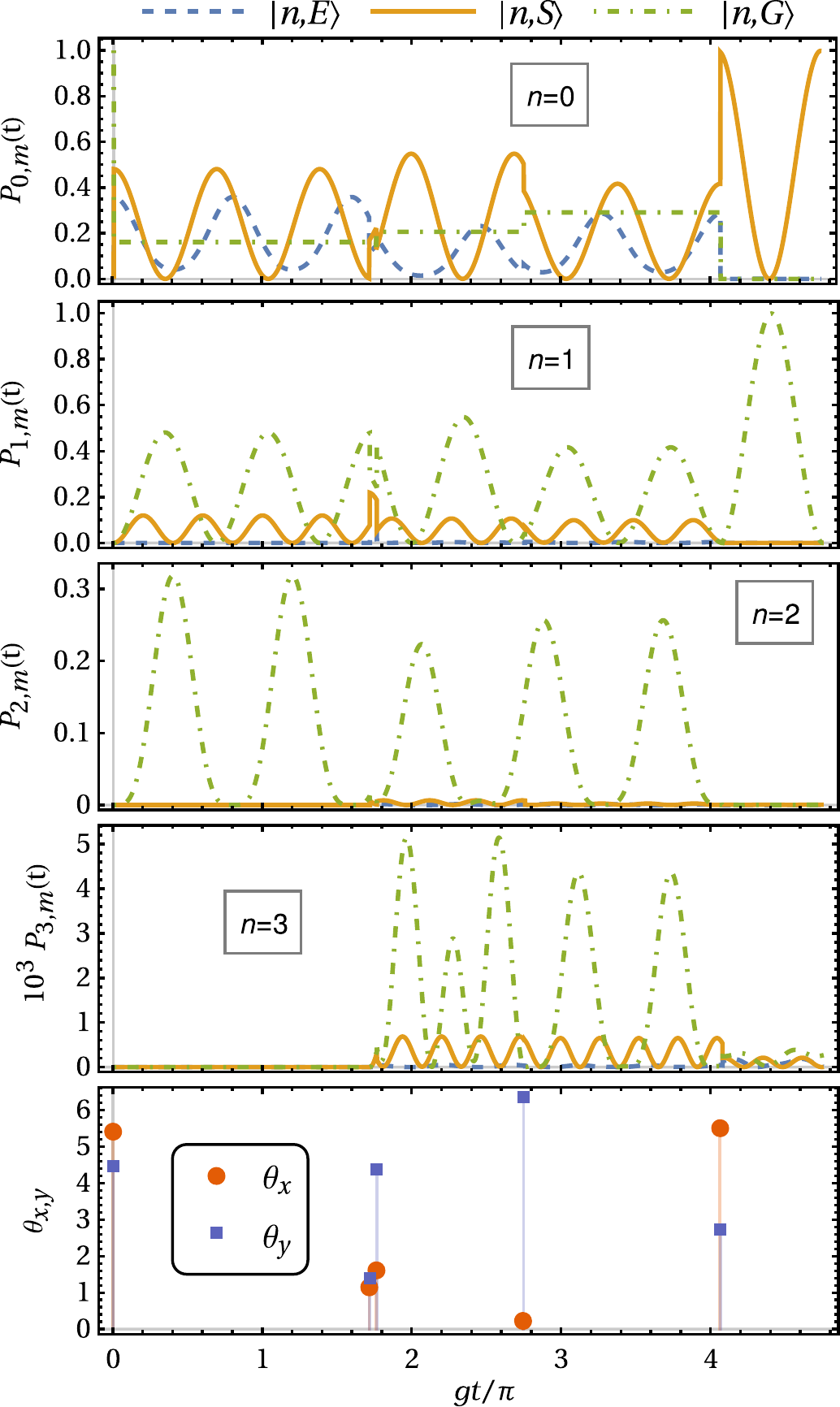}
\caption{Two-spin system without damping ($\kappa=0$) at resonance and without squeezing (its presence is negligible). Optimized dynamics in the case when the system is driven from $|0,G\rangle\langle 0,G|$ towards $|0,S\rangle\langle 0,S|$ by maximizing the fidelity with respect to this target state with the constraint $g t_{\mathrm{max}} =5 \pi$. The three upper panels show  the time evolution of the projection onto the states $\ket {n,m}$. The lower panel represents the pulse sequence. Note that each pulse induces a jump in the time evolution of the populations (dynamics are not continuous due to the presence of Dirac distributions in the control field). For simplicity, we consider for all the panels the case without squeezing. The numerical values of each $\theta_{x,y}$ are given in Tab.~\ref{tab:sequence_parameters}.}
\label{fig:traj_symmetric_5pi}
\end{figure}

We first investigate the generation of a symmetric state $\ket{0,S}$ with the corresponding density matrix $\hat \rho_{\mathrm{ts}}=|0,S\rangle\langle 0,S|$ called below the target state. The initial state is chosen to be $\hat \rho_{0}=|0,G\rangle\langle 0,G|$. We consider two spins at resonance ($\Delta_1=\Delta_2=0$), and we set $\kappa = 0$. We compute several pulse sequences that maximize the fidelity of Eq.~\eqref{eq:fidelity} with a time constraint $t_f \leq t_{\mathrm{max}}$ induced by fixing a maximal time $t_{\mathrm{max}}$, where we recall that $t_{f}$ is the control time. Numerical computations are performed using $x$ and $y$ collective controls, and with or without squeezing. The number of pulse packages is limited to seven. Results are presented in Fig.~\ref{fig:squeezing_speed_up} in which the  fidelity measures the efficiency of the control protocol. We observe that the squeezing slightly improves by a few percent the fidelity for short control times. In both cases, the target state is generated with good precision when $gt_{\mathrm{max}}\geq 2\pi$ ($gt_f \approx$ $1.99 \pi$). This duration corresponds to the shorter Rabi-frequency of the system. However, an almost perfect transfer ($F\simeq 1$) is only achieved for a very long control time. Here, $F\approx 0.998$ is obtained for $gt_{\mathrm{max}} = 5 \pi ~ (gt_f \approx4.85 \pi)$ by using 5 pulse packages with only coherent control pulses. Adding more packages or squeezing pulses does not improve $ F$ or does not reduce significantly $t_f$. On the basis of the very good fidelities achieved, we conjecture that the algorithm converges towards a state very close to the global optimum of the control process. The dynamics of the optimal solution and the pulse sequence for the case $g t_{\mathrm{max}} =5 \pi$ without squeezing are displayed in Fig.~\ref{fig:traj_symmetric_5pi}.

The control mechanism can be summarized as follows. The first and the last pulse packages play a key role. The first package excites the spins, while the last one allows us to reach approximatively a superposition of states of the form $\ket{\psi(t)}=A \cos (\sqrt{2}g t) \ket{1,G} + B \sin (\sqrt{2}g t) \ket{0,S} $. Pulse packages 2, 3, and 4 are designed by the optimization algorithm to achieve this goal approximatively. This trajectory can be obtained only with specific phase conditions between the states $\ket{0,E},\ket{0,S},\ket{1,G},\ket{1,S}$, and $\ket{2,G}$, which are connected together through $\hat{H}_0$ and $\hat{H}_C$. The relations are not trivial, because oscillation periods depend on the photon numbers, and these states could be dynamically connected to states with higher numbers of photons (such a transfer of population must be as small as possible).  Note that after the last pulse, the fidelity is not maximum, and almost one additional period (of a free evolution) is used to reach the true maximum. However, this is only slightly above the value after the last pulse. A similar process is observed  for shorter pulse sequences, but the positions and the amplitudes of the pulses are different. Note that the control protocol is not robust with respect to the number of spins because it depends strongly on Rabi oscillations. If the number of spins is changed, a new control field must be computed.

\begin{figure}
\includegraphics[width=\columnwidth]{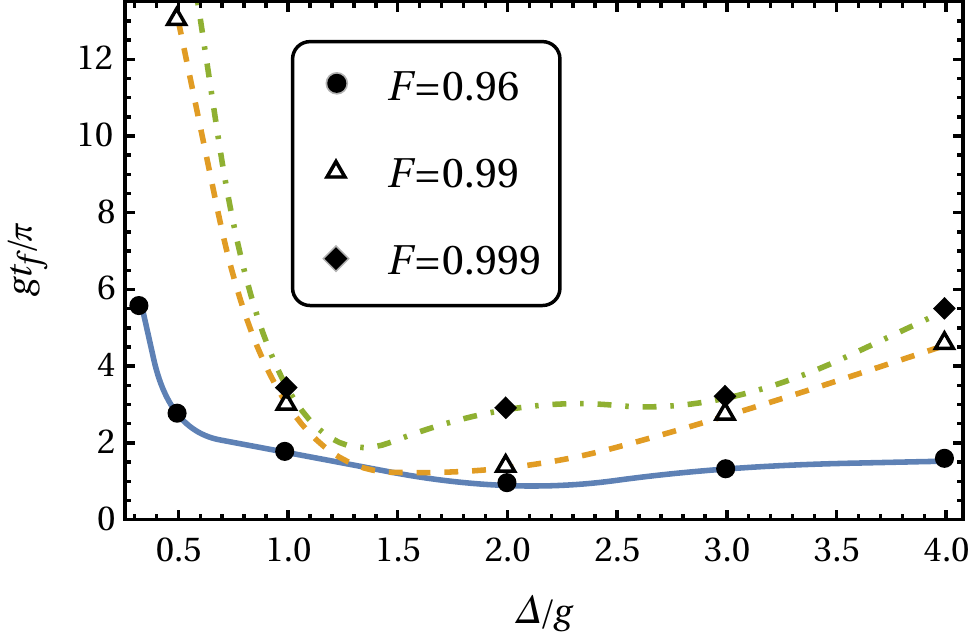}
\caption{Two-spin system without damping ($\kappa=0$) with $\Delta_1 = -\Delta$ and $\Delta_2 = \Delta$. The squeezing control is set to 0. Minimum time $gt_f/\pi$ to generate the antisymmetric state $\ket{0,A}$ as a function of the offset $\Delta/g$, for different values of fidelity. Smooth interpolation curves are plotted to guide the reading. Times above 13.5 are not represented (there is a point of the green dashed line at $gt_f/\pi = 16.8$ for $\Delta/g = 0.5$). Since the fidelity cannot be constrained precisely in the optimization algorithm, the points displayed in this figure are deduced from linear interpolation of the data set (for each point, three optimizations are made such that the desired fidelity is approximately achieved. Then, the value of $gt_f$ for a fixed value of $F$ is deduced from an interpolated curve). A pulse sequence with $F=0.99998$ with $\Delta/g = 1$ is given in Tab.~\ref{tab:sequence_parameters}.}
\label{fig:tf_fx_Delta}
\end{figure}

A similar study can be made to generate the state $\hat \rho_{\mathrm{ts}} =|0,A\rangle\langle 0,A|$. In this case, the spins are not taken in resonance in order to couple $\ket{0,A}$ with the other states.  We consider a symmetric distribution of offsets around the cavity frequency: $\Delta_1 = -\Delta$ and $\Delta_2 = \Delta$. Figure~\ref{fig:tf_fx_Delta} shows the evolution of the control duration as a function of  $\Delta/g$ for different values of the fidelity in the absence of the squeezing control. As could be expected, $F$ increases with the control time, which is minimized when $\Delta/g \in [1,2]$.
The latter result can be interpreted as follows. In the limit $\Delta \rightarrow 0$, the state $\ket{0,A}$ cannot be reached since it becomes dynamically disconnected from all the other states, which leads to $t_f \rightarrow \infty$. In the opposite case, the intermediate representation~\cite{messiah1962quantum} allows us to express the evolution operator between two pulses packages as:
\begin{equation}
\begin{split}
&e^{-\frac{\ii}{\hbar} t_k \hat H_0} = \underbrace{e^{- \ii\sum_n \frac{\Delta_n}{2} \sigz^{(n)} t_k}}_{\hat U_z(t_k)} \times \\
& \mathbb{T}\exp \left[- \ii\int_0^{t_k} \hat U_z ^{-1}(t) g\sum_{n=1}^{N_s} \left(\ad \sigm^{(n)}+\A \sigp^{(n)}\right)  \hat U_z(t) dt\right].
\end{split}
\end{equation}
The second time ordered exponential can be evaluated using a Magnus expansion. At first order, we obtain:
\begin{equation}
\begin{split}
&\int_0^{t_k} \hat U_z ^{-1}(t) g\sum_{n=1}^{N_s} \left(\ad \sigm^{(n)}+\A \sigp^{(n)}\right)  \hat U_z(t) dt = \\
& -\sum_{n=1}^{N_s} \frac{2 g \sin(\Delta_n t_k /2)}{\Delta_n} \left[ \ad \sigm^{(n)} e^{-\ii \Delta _n t_k/2}+\A \sigp^{(n)} e^{\ii \Delta _n t_k/2}\right].
\end{split}
\label{eq:first_order_magnus_with_detuning}
\end{equation}
We deduce that each spin contributes with an effective coupling proportional to $g/\Delta_n$. Similar calculations at higher orders introduce terms in power of $1/\Delta_n$. In the limit $|\Delta_n|\rightarrow \infty$, the above effective couplings to the cavity go to zero, and thus the spins cannot be entangled  in short times, since the interaction is the only mechanism that produces entanglement. This point explains why $t_f \rightarrow \infty$ when $\Delta \rightarrow \infty$.

\subsection{Optimal control in the presence of relaxation}

In this subsection, we consider again two spins at resonance and the target state $\hat \rho_{\mathrm{ts}} =\ket{0,S}\bra{0,S}$. The initial state is again $\hat \rho_{0}=|0,G\rangle\langle 0,G|$. We are interested in the robustness against the relaxation effect of the solutions derived at $\kappa =0$ and the possibility to limit the negative effect of the environment by a specific optimization of the control fields.
Similar results can be obtained for the antisymmetric state.

\begin{figure}
\includegraphics[width=\columnwidth]{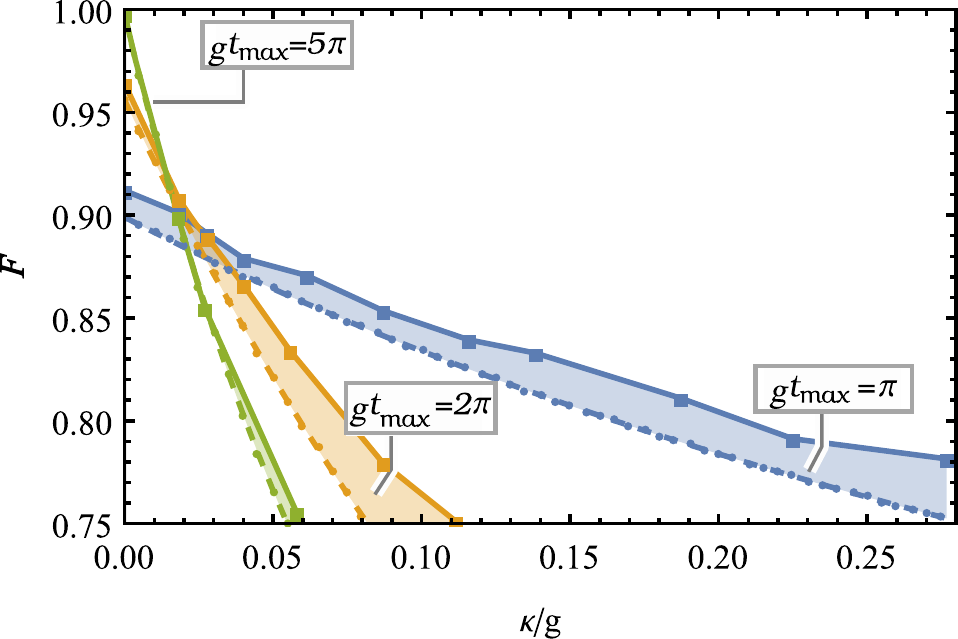}
\caption{Two-spin system with damping at resonance. Fidelity with respect to the symmetric state $\ket{S}$ as a function of $\kappa/g$ for the optimal control fields determined at $\kappa = 0$ (dashed lines) and the ones optimized in presence of relaxation (solid lines). In each case, squeezing control is also used. The squares and the dots indicate the points for which the computations have been done.}
\label{fig:fidelity_two_spins}
\end{figure}

Contrary to the case where $\kappa=0$, a brute force optimization is very arduous, and the algorithm is quickly trapped in a local optimum. To bypass this problem, optimizations taking into account the cavity damping are performed iteratively for increasing values of $\kappa$, starting from the solution at $\kappa=0$.
Optimizations are made for different values of $t_{\mathrm{max}}$, which is a fixed parameter in the algorithm. This is necessary because an initially long sequence does not converge in general toward a significantly shorter one, which may be more efficient. This is a consequence of the iterative optimization procedure that may return only a local maximum.

The robustness of the optimal solutions determined at $\kappa=0$ and the maximum fidelity obtained by numerical optimization for different values of $\kappa$ are plotted in Fig.~\ref{fig:fidelity_two_spins}. For a given value of $\kappa$, the best solution returned by the iterative re-optimization procedure  depends on the initial duration of the pulse sequence. In particular, for $\kappa/g \gtrsim 0.03$ short pulse sequences (e.g. with $gt_{\mathrm{max}} = \pi$) are more efficient than the longest ones (e.g. with $gt_{\mathrm{max}} = 5 \pi $). As in previous cases, the duration $t_f$  of a pulse sequence is slightly smaller than $t_{\mathrm{max}}$. If only coherent controls are used, then the results are qualitatively equivalent, the fidelity being 1 or 2 percents lower. Results in Fig.~\ref{fig:fidelity_two_spins} suggest the existence of an unreachable area, located above the  curves. This result shows a potential limit on the states that can be reached by the system. Although this observation is qualitatively expected~\cite{Altafini_2004}, we provide here a quantitative description of this effect. The numerical simulations shown in this figure reveal that the maximum fidelity decreases very quickly with $\kappa$, even with optimized controls fields.

\section{Optimization with a set of four spins}
\label{sec:many_spins}

In this section, we extend our study to a system composed of four spins.

\subsection{Optimal control without detuning and without relaxation}

\begin{figure}[t]
\includegraphics[width=\columnwidth]{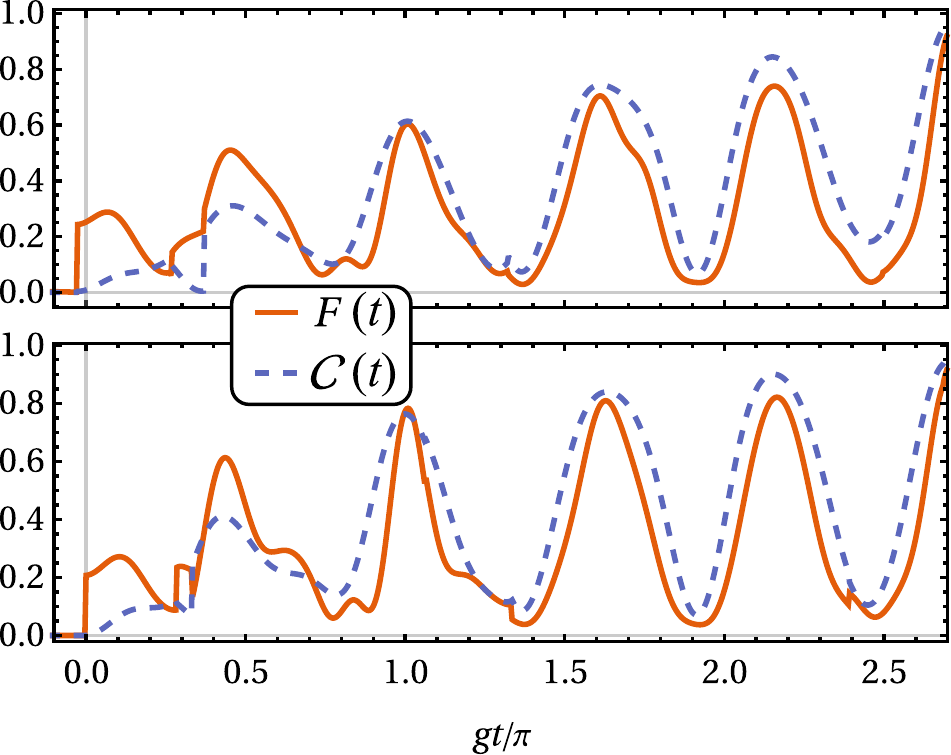}
\caption{Four-spin system without damping ($\kappa=0$) at resonance and without squeezing (its presence is negligible). Time evolution of the fidelity $F$ with respect to the state $\ket{S}$ and the cumulant-based quantity $\Mc C$ during the control process. The top and bottom panels represent respectively the dynamics driven by a control maximizing $\Mc C$ and $F$. The two pulse sequences do not have the same duration. To obtain the final states simultaneously, the first control starts before $t=0$. }
\label{fig:fidelity_cumulant_opt}
\end{figure}

We start the analysis by considering a system of four spins at resonance in a lossless cavity ($\Delta_n = 0$ and $\kappa =0$). As in the previous cases, the initial state is $\hat \rho_{0}=|0,G\rangle\langle 0,G|$. We compare optimizations concerning the maximization of the fidelity $F$ with respect to the state $\ket{S}$ and the cumulant-based quantity $\Mc C$ of Eq.~\eqref{eq:def_cumulant_FoM}. We stress that the fidelity is computed after tracing out the cavity degrees of freedom in order to provide a fair comparison between the optimizations ($\Mc C$ depends only on spin-operators). After maximization of $F$, we have obtained the following values at the final time:  $\Mc C \approx 0.95$, $F \approx 0.93$, and $gt_f \approx 2.70 \pi$. After maximization of $\Mc C$, similar values are achieved for $\Mc C$ at the final time: $\Mc C \approx 0.96$, $F \approx 0.92$, and $gt_f \approx 2.87 \pi$. In both cases, the time evolution of $F$ and $\Mc C$ during the control process are plotted in Fig.~\ref{fig:fidelity_cumulant_opt}. The two optimizations are performed using the same pre-optimization involving five pulse packages and leading to a fidelity of the pre-optimization $F \approx 0.84$. Final optimizations use seven pulse packages. Additionally, the time constraint is $gt_{\mathrm{max}}= 3 \pi$.
The two figures of merit provide similar results after optimization. As in the two-spin case for large values of $F$, in the considered situation, the presence of the squeezing control is negligible in the process.
\subsection{Optimal control with offset and relaxation}

\label{sec:Optimization with offset and relaxation}

The ability of generating the four-spin symmetric entangled state $\ket{S}$ decreases considerably in the case when the spins are not at resonance with the cavity mode. Here, we consider the following distribution: $\Delta_1/g = -1$, $\Delta_2/g = -0.5$, $\Delta_3/g = 0.5$, and $\Delta_4/g =1$. Again, the initial state is $\hat \rho_{0}=|0,G\rangle\langle 0,G|$. In the case without relaxation and with $gt_{\mathrm{max}}= 3 \pi$, by using $\Mc C$ as figure of merit to maximize we have not been able to find a better value than $\Mc C \approx 0.73$. We interpret this result as a bad convergence of the algorithm. In principle, a better fidelity should be achieved with a longer control time, and more pulse packages. Unfortunately, the numerical cost for such optimization process is unreasonable. Similar difficulties have been encountered for maximizing $-\Mc C$ (i.e., aiming to generate antisymmetric states). These problems are not specific to the considered figure of merit. Indeed, they have been encountered with other quantifiers as well, for which the results have been worse (see Appendix \ref{sec:state_properties_relaxation}, for the list of the other quantifiers we have considered). As in the case of Sec.~\ref{sec:2_spins}, we observe that when $\kappa >0$, the time $t_{\mathrm{max}}$ can be reduced to improve the algorithm convergence.

Contrary to the cases of Sec.~\ref{sec:2_spins} where precise target states of simple form have been used, here the states generated by the optimization process are quite complex superradiant states. Moreover, the spins and the cavity are in a non-classical state. We refer to Appendix~\ref{sec:state_properties_relaxation} for further details on the state characterization. In order to visualize in a simple way the gain of non-classicality produced by the control fields, the optimized $\Mc C (t_f)$ is compared to the maximum value of the same quantity obtained with the free dynamics, $\Mc C_{\mathrm{max}} = \max_{t\in [0, t_{\mathrm{max}}]} \Mc C (t)$. The free evolutions are initialized with a $\pi$-pulse, to generate the excited state $\hat \rho = \ket{0,E}\bra{0,E}$. The reference dynamic is similar to the one used in superradiance experiments, where the spins decay through the superradiant subspace of the Hilbert space. In the bad-cavity limit, this leads to a superradiant emission of photons. Additional elements of comparison with state-of-the-art physical quantities are given in Tab.~\ref{tab:cumulant_tab} of Appendix~\ref{sec:state_properties_relaxation}.

\begin{figure}[t]
\includegraphics[width=\columnwidth]{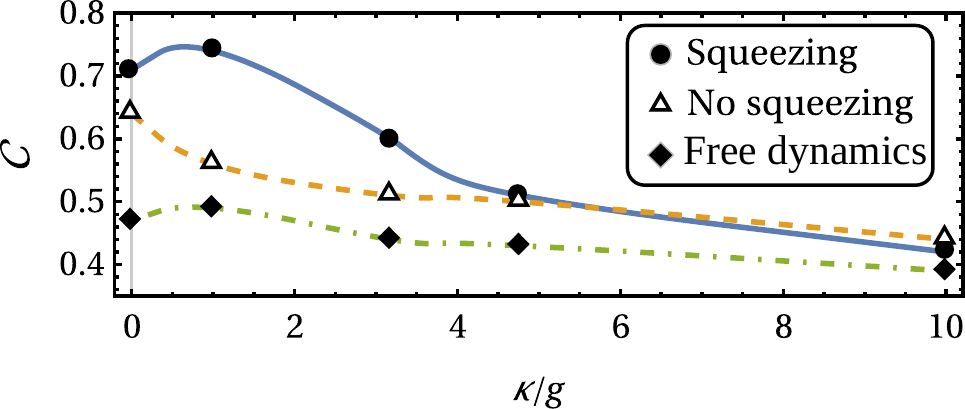}
\caption{Four-spin system with damping with $\Delta_1/g = -1$, $\Delta_2/g = -0.5$, $\Delta_3/g = 0.5$, and $\Delta_4/g =1$. Maximum of $\Mc C$ after optimization with or without squeezing with respect to $\kappa/g$. In the case of a $\pi$-pulse, the values are the maximum in the interval $[0,g t_{\mathrm{max}}]$, with $g t_{\mathrm{max}}=\pi/2$. Interpolated curves are used to guide the reading. The pulse sequence used to generate the point with $\kappa/g =1$ (with squeezing) is given in Tab.~\ref{tab:sequence_parameters}.}
\label{fig:Cumulant_fx_kappa}
\end{figure}
Figure~\ref{fig:Cumulant_fx_kappa} shows the value of $\Mc C$ as a function of $\kappa/g$, which are obtained for different optimized control fields, and for the free dynamics.
We observe a non-negligible enhancement of the entanglement process with squeezing controls for small values of $\kappa$ ($\kappa/g <5$). This advantage is not present when the system is close to the bad-cavity regime ($\kappa/g = 10$). In this latter case, it seems difficult to obtain states that are significantly more entangled than the ones naturally obtained by free dynamics. Notice that this happens far from the semi-classical regime, which starts from $\Mc C \simeq 0.15$ for the system studied in this section. Interestingly, the gain offered by squeezing fields seems related to the cooperativity parameter~\cite{temnov_superradiance_2005}
\begin{equation}
\mathsf{C} = \frac{2 g^2 N_s}{\kappa \Omega},
\end{equation}
where $\Omega$ is the full width of the offset distribution. Cooperative effects are dominant when $\mathsf{C}\gg 1$ while the semi-classical regime is obtained when $\mathsf{C} \ll 1$. The boundary between the two regimes can be estimated using $\Omega /g = 2$. Then, the cooperativity is equal to 1 for $\kappa/g = 4$, which is close to the point in Fig.~\ref{fig:Cumulant_fx_kappa} where the squeezing starts to offer an advantage. It seems that squeezing fields are interesting only when cooperative effects play a major role in the system dynamics, although a deeper analysis is required to validate this first observation.

We finish this section by highlighting the fact that a relaxation process can be a (limited) vector of entanglement~\cite{choi_quantum_2007}. This point is clearly visible in Fig.~\ref{fig:Cumulant_fx_kappa} (for the free dynamics, the maximum value of $\Mc C$ is obtained for $\kappa/g = 1$, and not $\kappa/g = 0$). A simple explanation is given by considering a spin ensemble initially in the state $\ket{0,E}$. Without relaxation, the system cannot visit the state $\ket{S}$, whereas in the presence of damping this state can be partially populated without external action on the system.

\section{Conclusion}
\label{sec:conclusion}

In this paper, we have studied the generation of non-classical states in a system of spins coupled to a cavity. We have shown that a series of short coherent and squeezing controls allows in some cases to reach such states with a good fidelity or with large values of a cumulant-based measure. A specific numerical optimization procedure has been developed to find the parameters of the pulse sequences. The proposed control strategy has also the decisive advantage of being experimentally implementable.

The physical limit of the different control objectives has been studied as a function of the cavity damping parameter and spin offsets, for several configurations of two and four spins. In general, the detrimental effect induced by the environmental noise cannot be removed efficiently by the pulse sequence. Based on numerical evidence, we conjecture the existence of a large unreachable area in the space of density matrices. The presence of spin offset is required to couple antisymmetric subspaces with other states, but it also makes the spin dynamics harder to control. For large offsets, optimal control has limited performances. In parallel, we have studied the gain achieved by adding squeezing field to the pulse sequence. While in general, squeezing enables only a slight enhancement of the efficiency of the control process, a huge improvement is observed in the case of the maximization of the cumulant-based figure of merit in the good cavity regime for a four-spin system out of resonance.

This study opens the way to many challenging theoretical and experimental issues. A possible perspective consists in carrying out the same type of analysis but with a complex non-Markovian environment replacing the current Markovian bath of the cavity. It has been recently shown that optimal control algorithms can be applied in this setting~\cite{koch_controlling_2016,wilhelm2009,mangaud2018,reich}. Different works have proved that non-Markovianity (or at least employing a structured environment) could be used as a valuable resource for quantum control, by allowing new and efficient control mechanisms that are not feasible with Markovian dynamics.

\section*{Acknowledgments}

This work was mainly supported by the French ``Investissements d'Avenir'' program, Project ISITE-BFC (Contract No.~ANR-15-IDEX-03). This research has been partially supported by the ANR projects COQS (Contract No.~ANR-15-CE30-0023-01) and QUACO (Contract No.~ANR-17-CE40-0007-01).
This project has also received funding from the European Union Horizon 2020 research and innovation program under the Marie Skłodowska-Curie Grant No.~765267 (QUSCO).

\appendix

\section{Properties of the states generated by the maximization of $\Mc C$}

\label{sec:state_properties_relaxation}

We present in this section a short analysis of the several states produced by the maximization of the cumulant-based quantity of Eq.~\eqref{eq:def_cumulant_FoM}. As emphasized in Sec.~\ref{sec:many_spins}, the final states are in a complex superposition of entangled states. In order to highlight some of their physical properties, we compute the following quantities.
\begin{itemize}
	\item The average dipole-dipole operator $\langle \hat J_+ \hat J_-\rangle$. For a four-spin system, the state $\ket{n,E}$ returns the value 4 and $\ket{n,S}$ returns 6 (the maximum).
	\item The dipole-dipole correlator $\langle \hat J_+ \hat J_-\rangle_{\mathrm{corr}}= \langle \hat J_+ \hat J_-\rangle - \sum_n \langle \sigp^{(n)} \sigm^{(n)}\rangle$. For a four-spin system, the state $\ket{n,E}$ returns the value 0 and $\ket{n,S}$ returns 4 (the maximum).
	\item The second order correlation function of the cavity field  $g^{(2)}= \langle \ad \ad \A \A\rangle/\langle \ad \A \rangle^2$~\cite{ficek_entangled_2002,auffeves_few_2011}. It takes the value 1 if the field is in a coherent state.
\end{itemize}

The values of these quantities for the final states determined in Sec.~\ref{sec:Optimization with offset and relaxation} for the cases $\kappa/g = 0$, 1, and 10, are given in Tab.~\ref{tab:cumulant_tab}. The maximum values obtained with free dynamics are also included in this table.
\begin{table}[t]
	\begin{center}
		\begin{small}
			\begin{tabular}{c|c|cccc} \toprule
				&$\kappa/g$     & 0        & 1      & 10     \\ \midrule
				& Cooperativity $\mathsf{C}$ &$\infty$  & 4      & 0.4    \\ \midrule
				\multirow{2}{7em}{\centering
					Optimized control with squeezing
				}                             &$\Mc C$                                   & 0.71 & 0.74  & 0.42   \\
				&$\langle \hat J_+ \hat J_-\rangle_{\mathrm{norm}}$        & 0.83 & 0.87  & 0.78  \\
				&$\langle \hat J_+ \hat J_-\rangle_{\mathrm{corr, norm}}$ & 0.73 & 0.76  & 0.45   \\
				&$g^{(2)}$                                 & 1.87 & 1.84  & 2.17   \\
				\midrule
				\multirow{2}{7em}{\centering
					Optimized control without squeezing
				}                             &$\Mc C$                                   & 0.64 & 0.56  & 0.44   \\
				&$\langle \hat J_+ \hat J_-\rangle_{\mathrm{norm}}$        & 0.70 & 0.72  & 0.74  \\
				&$\langle \hat J_+ \hat J_-\rangle_{\mathrm{corr, norm}}$ & 0.64 & 0.62  & 0.45  \\
				&$g^{(2)}$                                 & 1.81 & 1.67  & 1.45  \\
				\midrule
				
				\multirow{2}{6em}{\centering
					Free dynamics
				}                             &$\max_t \Mc C$                                 & 0.47 & 0.49 & 0.39    \\
				&$\max_t\langle \hat J_+ \hat J_-\rangle_{\mathrm{norm}}$       & 0.78 & 0.79  & 0.78   \\
				&$\max_t\langle \hat J_+ \hat J_-\rangle_{\mathrm{corr, norm}}$& 0.47 & 0.49  & 0.38  \\
				&$\max_t g^{(2)}$                               & 1.50 & 1.50  & 2.7  \\
				\midrule
			\end{tabular}
		\end{small}
	\end{center}
	\caption{Values of $\Mc C$ and other physical quantities of the final state after optimization. $\langle \hat J_+ \hat J_-\rangle$ and $\langle \hat J_+ \hat J_-\rangle_{\mathrm{corr}}$ are normalized by their maximum values. The optimizations are performed using different $\kappa$ and $gt_{\mathrm{max}} = \pi/2$. We compare optimizations with the maximum values obtained with the free dynamics of a spin ensemble initially in the state $\hat \rho = \ket{0,E}\bra{0,E}$. The pulse sequence used to generate the data with $\kappa/g =1$ with squeezing is given in Tab.~\ref{tab:sequence_parameters}. }
	\label{tab:cumulant_tab}
\end{table}

We observe that the final states belong to the superradiant subspace since $\langle \hat J_+ \hat J_-\rangle_{\mathrm{corr}}>0$ ~\cite{temnov_superradiance_2005,mascarenhas_cooperativity_2013}. In particular, the threshold $\langle \hat J_+ \hat J_-\rangle_{\mathrm{norm}}\equiv\langle \hat J_+ \hat J_-\rangle/6=N_s/6 \approx 0.67$ is exceeded. This is interesting since in the bad-cavity limit this means that the emission rate is larger than the emission rate of $N_s$ independent excited spins (since in this limit $\langle \hat J_+ \hat J_-\rangle$ is proportional to the spin-ensemble emission rate)~\cite{gegg_superradiant_2018,mascarenhas_cooperativity_2013,andreev_collective_1980,gross_superradiance:_1982}. In general, this is not true and the emission rate depends on the number of photons contained in the cavity. Therefore, this kind of interpretation cannot be performed here since we are not in the bad-cavity regime. In this sense, exceeding this threshold value remains just an indication of the superadiant collective behaviour of the spins. Concerning the nature  of the cavity state, it is described by $g^{(2)}$. For all cases, we have $g^{(2)} $ between $1.45$ and $2.7$. Hence, the cavity state has a non-classical photon statistic, which is  induced by the process of building up of entanglement with the spin ensemble and by the squeezing control field. The resulting photon emission from the spin ensemble is very different from the situation encountered in the bad-cavity limit, where the cavity field is close to a coherent state ($g^{(2)}=1$).

We finish this section with another observation. In the case $\kappa/g = 10$ the values of all the quantities, $\Mc C$, $\langle \hat J_+ \hat J_-\rangle$, and $\langle \hat J_+ \hat J_-\rangle_{\mathrm{corr}}$, are similar for all the different final states. This supports our conclusion regarding the limited effect of optimal control in the weak cooperativity regime (see Sec.~\ref{sec:Optimization with offset and relaxation}).

\section{The short pulse limit and the squeezing control}
\label{sec:appendix The short pulse limit and the squeezing control}

This appendix is dedicated to technical details regarding the control fields. First, we briefly recall the short pulse approximation and its realization using bump pulses. Then, we present the physical effect of squeezing control on spin dynamics.

\subsection{The short pulse limit}
\label{sec:short_pulse_limit}

We introduce below the bump pulses in the short pulse limit. A complete and detailed presentation is given in Refs.~\cite{ansel_optimal_2018,ansel_optimal_2018-1}. For an experimental implementation, see Ref.~\cite{probst_shaped_2019}.

For pedagogical reasons, we introduce the approximation by means of observable expectation values, and we restrict the control to the $x$ axis, i.e., with $\coe \in \setR$. The generalization to $x$ and $y$ controls ($\coe \in \setC$), and with operators in the Heisenberg or in the Schr\"odinger picture is possible. Furthermore, we assume that $\squ=0$, and for simplicity $\Delta_n = 0$ $\forall n$. We recall that in Eq.~\eqref{eq:full_Hamiltonian} the amplitude of $\coe$ must be smaller than the cavity frequency in order to verify the fixed dissipator approximation~\cite{giorgi_microscopic_2020}.
The underlying idea consists of determining the effect of the control field on the spin system, and to find a solution that verifies a list of properties in the limit when the control duration is very small with respect to the spin-cavity characteristic interaction time.

For that purpose, we need to compute the time evolution of the electromagnetic field inside the cavity. We introduce the quadratures $ X = \langle \ad + \A \rangle$ and $ Y = -\ii\langle \ad - \A \rangle$. The differential equation governing their evolution is determined from the relation $d_t O = \text{Tr}[\hat O d_t \hat \rho]$, where the average of  an arbitrary operator $\hat O$ is indicated by $O= \text{Tr}[\hat O  \hat \rho]$. After the formal  integration of $d_t X$, one obtains:
\begin{equation}
 X(t) = \int_0^t e^{-\kappa (t-t')/2} \left[\coe (t') - g  J_y(t') \right]dt' +  X(0).
\label{eq:solution_X_quadrature}
\end{equation}
A similar expression can be established as well for $Y(t)$, but we do not need it explicitly. Following the same direction, we derive the Bloch equations:
\begin{equation}
\frac{d}{dt} \left( \begin{array}{c}
 J_x (t) \\
 J_y (t) \\
 J_z (t)
\end{array} \right)= g\left( \begin{array}{c}
 X(t) \\
 Y(t) \\
0
\end{array} \right) \wedge \left( \begin{array}{c}
 J_x (t) \\
 J_y (t) \\
 J_z (t)
\end{array} \right),
\end{equation}
where $\wedge$ is the vector product. In the limit $T\ll 1/g$, where $T$ is the pulse duration, the evolution of the cavity field follows the dynamics of the drive and the back reaction of the spin ensemble is negligible (first order in $g$). Therefore, we have:
\begin{equation}
\frac{d}{dt} \left( \begin{array}{c}
 J_x  \\
 J_y \\
 J_z
\end{array} \right) \approx g\left( \begin{array}{c}
 X(0) + A(t) \\
 Y(0) \\
0
\end{array} \right) \wedge \left( \begin{array}{c}
 J_x  \\
 J_y  \\
 J_z
\end{array} \right),
\label{eq:approximated_bloch_equation_1}
\end{equation}
where we have introduced the function
\begin{equation}
A(t)=\int_0^t e^{-\kappa (t-t')/2}\coe (t')dt'.
\end{equation}
We point out that if we construct $\coe$ so that:
\begin{equation}
\left\lbrace \begin{array}{cc}
A(T) & = 0 \\
\int_0^T A(t) dt & =  \theta/g
\end{array} \right. ,
\label{eq:conditions_bump_pulse}
\end{equation}
we obtain an approximation of a Dirac distribution $\delta_0(t)$. Inserting \eqref{eq:conditions_bump_pulse} into \eqref{eq:solution_X_quadrature} and \eqref{eq:approximated_bloch_equation_1} leads to:
\begin{equation}
\begin{split}
 X(T) & \approx  X(0) \\
\frac{d}{dt} \left( \begin{array}{c}
 J_x  \\
 J_y  \\
 J_z
\end{array} \right)& \approx g\left( \begin{array}{c}
 X(0) + \frac{\theta}{g} \delta_0(t) \\
 Y(0) \\
0
\end{array} \right) \wedge \left( \begin{array}{c}
 J_x  \\
 J_y  \\
 J_z
\end{array} \right).
\end{split}
\end{equation}

A very simple solution of this system is given by bump pulses or other optimized generalizations. A bump pulse is parameterized by the following function:
\begin{equation}
\coe(t) = \frac{2\theta C_p}{g T}  \left[\frac{\kappa}{2}-\frac{T^2(T-2t)}{4 t^2 (t-T)^2}\right]e^{\frac{T^2}{4t^2 - 4tT}}  \mathbb{I}_{[0,T]}(t),
\label{eq:bump_pulse}
	\end{equation}
where $\mathbb{I}_I$ is the indicator function on the interval $I$ (this function takes the value 1 for elements of $I$ and 0 outside) and the constant $C_p$ is chosen so that the Eq.~\eqref{eq:conditions_bump_pulse} is verified. It can be expressed in terms of Whittaker's function~\cite{magnus2013formulas}: $C_p= \sqrt{\pi/e}W_{-1/2,1/2}(1) = 0.44399\cdots $.

\subsection{Squeezing control}
\label{sec:squeezing control}

We present here the physical effect of a squeezing control on the spin dynamics. Similar calculations can be found in Refs.~\cite{gardiner_driving_1994,gardiner_inhibition_1986,zeytinoglu_engineering_2017}.

We analyze the dynamics in the interaction picture. For simplicity, we assume $\kappa = 0$ (a generalization is straightforward) and we introduce the evolution operator $\hat U[\hat O]$, solution of $d_t \hat U[\hat O] = -\frac{\ii}{\hbar} \hat O \hat U[\hat O]$. The intermediate representation theorem \cite{messiah1962quantum} gives:
\begin{equation}
\hat U[\hat H] = \hat U_s \hat U[\hat U_s^{-1}\hat H' \hat U_s],
\end{equation}
where $\hat U_s = \hat U [\ii  \hbar s ((\ad)^2 -(\A)^2 )/2]$, $s = \int_0^t \squ(t')dt'$, and $\hat H' = \hat H - \ii \hbar \squ (t')[(\ad)^2 -(\A)^2]/2$. In the rotating frame, the squeezing generator produces a Bogoliubov transformation of creation and annihilation operators:
\begin{equation}
\begin{split}
\hat U_s^{-1} \A \hat U_s &=  \text{ch}(s)\A +  \text{sh}(s) \ad \equiv \hat b \\
\hat U_s ^{-1} \ad \hat U_s &= \text{ch}(s)\ad + \text{sh}(s) \A \equiv \hat b^\dagger .
\end{split}
\end{equation}
The new expression of the Lindblad equation is obtained by replacing $\A \rightarrow \hat b (s(t))$, and $\ad \rightarrow \hat b^\dagger (s(t))$. Hence, all the terms of $\hat H'$ depend non-trivially on the squeezing.
In the limit $ s\rightarrow 0$, we have $\A = \hat b $, while in the limit $ s \rightarrow \pm \infty$, we obtain $\hat b = e^{s}(\A \pm \ad)/2$. In the $+\infty$ limit, the interaction Hamiltonian becomes approximatively:
\begin{equation}
\frac{ \hbar g}{2}e^s(\ad + \A )\hat J_{x}.
\end{equation}
Therefore, the squeezing enhances the coupling strength by a factor $e^s/2$, but it also changes the interaction into a spin-boson interaction. Using $s=2$ (upper bound allowed in numerical optimizations), we get $\text{ch} (s) \approx 3.76$,  $\text{sh} (s)  \approx 3.62$, and $e^s/2 \approx 3.69$.

Finally, we emphasize that it is possible to show that the opposite effect can be obtained by setting $\beta \rightarrow \beta e^{\ii \pi/2}$. We refer to Ref.~\cite{nieto_holstein-primakoff/bogoliubov_1997} for a precise presentation of Bogoliubov transformations in the complex case. In this case, the effective coupling decreases. For this reason, we have not considered in this paper the general setting $\beta \in \setC$.

\section{Pulse sequence parameters, robustness, and numerical optimization procedure}
\label{sec:sequence_parameters}

\begin{table*}[t]
	\begin{small}
		\begin{enumerate}
			\item \textit{Symmetric state:} two spins, $\kappa =0$, $\Delta_1 = \Delta_2 = 0$, $F=0.998$. \\
			\begin{tabular}{c|cccccc} \toprule
				$k$  & $\theta_{k,x}$ & $\theta_{k,x}$ & $ s_k$ & $gt_k$ \\ \midrule
				1  & 5.42           & 4.41           & 0      & 5.51  \\ \midrule
				2  & 1.16           & 1.34           & 0      & 0.15  \\ \midrule
				3  & 1.62           & 4.33           & 0      & 3.18  \\ \midrule
				4  & 0.23           & 6.28           & 0      & 4.24  \\ \midrule
				5  & 5.52           & 2.68           & 0      & 2.17  \\ \bottomrule
			\end{tabular}
			\raisebox{-.5\height}{\includegraphics[width=13.8cm,]{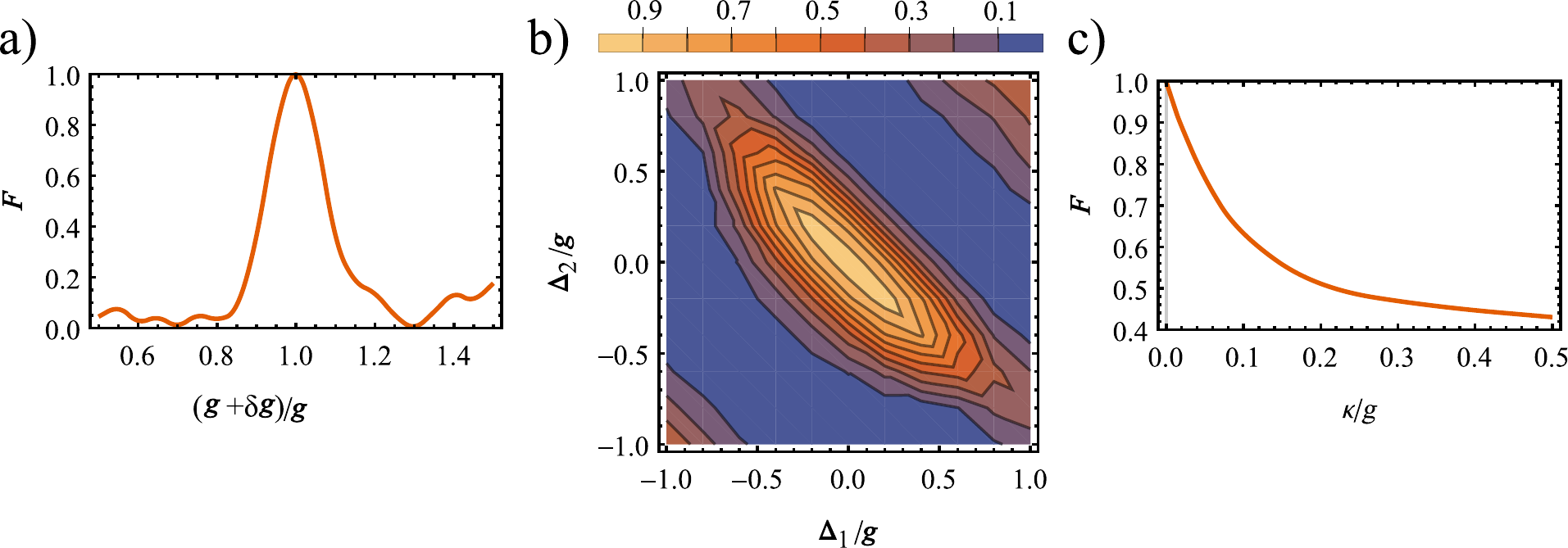}}
			\hspace{0.5cm}
			
			\item\textit{ Antisymmetric state:} two spins, $\kappa =0$, $ \Delta/g = 1$, $F=0.99998$.\\
			\begin{tabular}{c|cccccc} \toprule
				$k$  & $\theta_{k,x}$ & $\theta_{k,x}$ & $ s_k$ & $gt_k$ \\ \midrule
				1  & 2.49           & 4.69           & 0      & 2.40  \\ \midrule
				2  & 3.06           & 3.16           & 0      & 4.83  \\ \midrule
				3  & 1.62           & 5.61           & 0      & 2.38  \\ \midrule
				4  & 4.77           & 1.57           & 0      & 0.00  \\ \midrule
				5  & 4.74           & 3.16           & 0      & 1.21  \\ \bottomrule
			\end{tabular}
						\raisebox{-.5\height}{\includegraphics[width=13.8cm,]{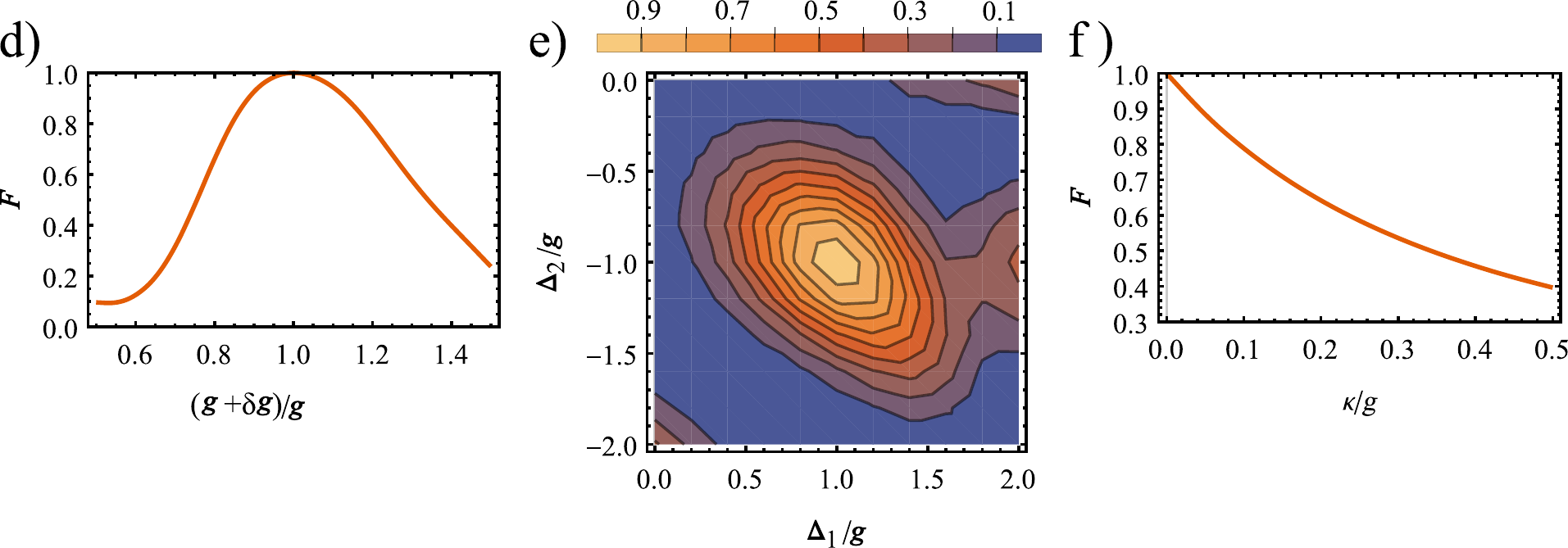}}
			\item \textit{Non-classical state:} four spins, $\kappa/g =1$, $s_{\mathrm{max}}=0.5$, $\Delta_1/g = -1$, $\Delta_2/g = -0.5$, $\Delta_3/g = 0.5$, $\Delta_4/g =1$, and $\Mc C=0.74$.\\
			\begin{tabular}{c|cccccc} \toprule
				$k$  & $\theta_{k,x}$ & $\theta_{k,x}$ & $ s_k$ & $gt_k$ \\ \midrule
				1  & 3.14           & 0.36            & 0.5      & 1.17  \\ \midrule
				2  & 3.14           & -3.14           & 0.5      & 0.40  \\ \midrule
				3  & 3.14           & -2.82           & 0.5      & 0.01  \\ \midrule
				4  & -3.14          & 2.84            & 0.5      & 0.63  \\ \midrule
				5  & 1.55           & 1.66            & 0.5      & 0.35  \\ \midrule
				6  & 3.14           & 3.14            & -0.06    & 0.00  \\ \midrule
				7  & 0.99           & -3.02           & 0.5      & 0.00 \\ \bottomrule
			\end{tabular}
			\raisebox{-.6\height}{\includegraphics[width=13.8cm,]{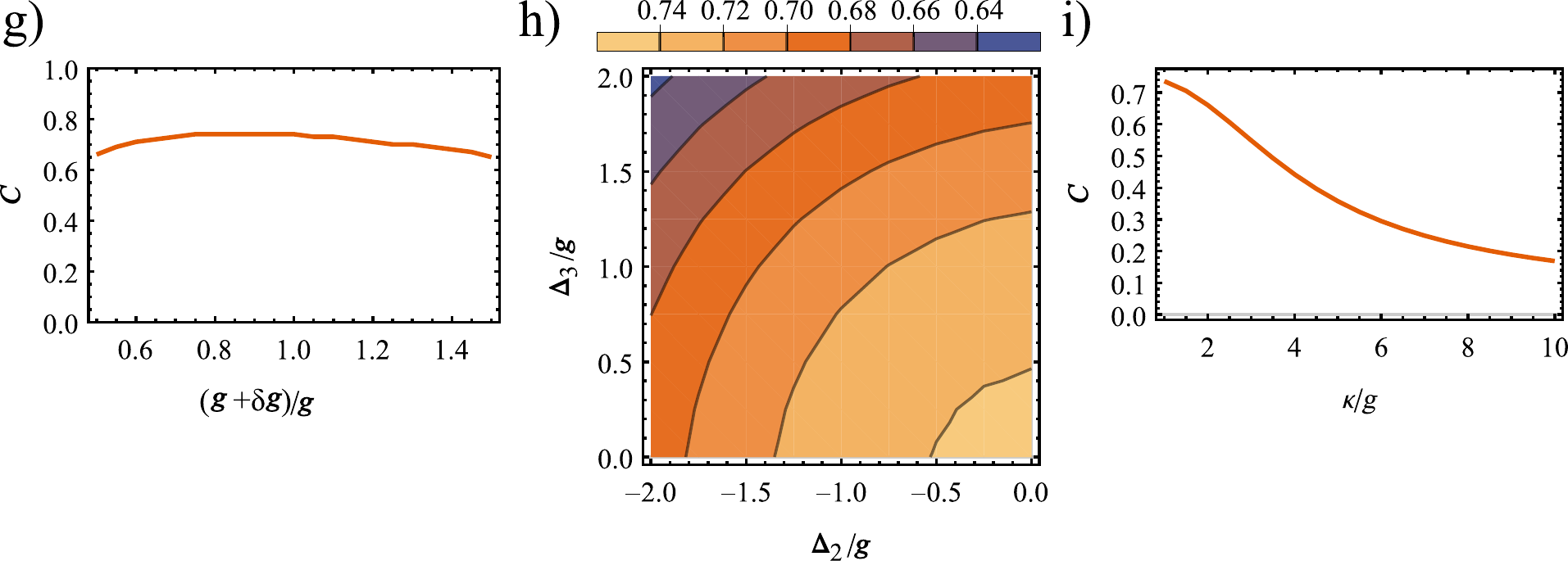}}
		\end{enumerate}
	\end{small}
	\caption{Parameters of some optimized pulse sequences and their robustness against some parameter variations (coupling strength $g$, offsets $\Delta_i$, and cavity damping $\kappa$). The sequence 1 is used to compute the point at $gt_{\text{max}}=5\pi$ (no squeezing) of Fig.~\ref{fig:squeezing_speed_up}. It is also the control field presented in Fig.~\ref{fig:traj_symmetric_5pi}. The sequence 2 is used to compute the point $\Delta/g = 1$ and $F=0.99998$ in Fig.~\ref{fig:tf_fx_Delta}. The sequence 3 is used to compute the point $\kappa/g =1$ with squeezing of Tab.~\ref{tab:cumulant_tab} and Fig.~\ref{fig:Cumulant_fx_kappa}. For all panels a), b), ..., i), the quantities which are not varied are kept fixed to the value used during the optimization process. For instance, in panels a), d), and g), the coupling strength is changed from the reference value $g$, the variation being denoted $\delta g$. In graphs b) and e), the color map represents the value of the fidelity $F$, while in h) it depicts the value of $\Mc C$. For the third pulse sequence, the offset distribution depends on four parameters. To simplify the analysis, we have only modified the offsets $\Delta_2$ and $\Delta_3$.}
	\label{tab:sequence_parameters}
\end{table*}
The parameters of a few control pulses used in this study are reported in Tab.$~$\ref{tab:sequence_parameters}. We have also included in the table several plots showing the robustness  of the optimal control against variations of the system parameters. The robustness of a pulse sequence relies on its ability to maintain a high fidelity of the control process when the system parameters are varied with respect to the values used to derive this optimized control sequence. This aspect is crucial for experimental implementations in which the parameters are only known to a given precision. Interestingly, numerical simulations reveal that the control used to generate the antisymmetric state is more robust than the one used to generate the symmetric state. Interestingly, the third control of the table, which maximizes $\Mc C$, is very robust against variations of all parameters (in particular, the value of $\Mc C$ may even increase).

We now briefly describe the numerical optimization procedure in which a two-step protocol has been used. First, a Gradient Ascent Algorithm (GAA), given by the \emph{NMaximize} function of \textit{Mathematica}, has been applied to pre-optimize a simplified system, i.e., a state to state transfer problem without decoherence and squeezing terms and with a maximum of 5 pulse packages. This first step is crucial for the second optimization which aims at computing a control field for  the full system using more pulse packages. For that purpose, a home-made version of the JAYA algorithm~\cite{venkata_rao_jaya:_2016} has been developed.

The JAYA algorithm is a gradient-free optimization algorithm that supports parallel computation (up to 8 cores in this study). It is a population-based method which repeatedly modifies a population of individual solutions, like genetic or simulated annealing algorithms. Compared to these common methods JAYA does not contain any free hyperparameters (such as the effective temperature in simulated annealing)~\cite{venkata_rao_jaya:_2016}. 
At a given iteration, the jump probability of a particle in the control landscape is entirely determined by the value of the cost function associated with all other particles. Then, particles are attracted toward the best solutions, and at the same time they avoid the worst solutions. The efficiency of the algorithm depends only on the number of particles and the number of iterations. A fine adjustment of these two parameters allows one to reduce the numerical cost of the optimization.

 The second optimization process is initialized with a set of control fields uniformly sampled around the pre-optimization solution (the distribution is of the order of $10\%$ around the initial value). The typical dimension of the system varies between 33 and 16384. The optimization computation time goes from 30 min to 3 days. A single propagation of the whole dynamics is quite fast, between $0.01$~s and a few seconds. However, the required number of iterations can be very large. With the JAYA algorithm, this number can be reduced to $5 \times 10^5$, but with a GAA, the number of iterations is larger than $5 \times 10^6$. More than 30~GB of RAM has been used to save data  and to speed up the optimization process.
Note that the two algorithms are global optimization algorithms in the sense that they use several initial conditions (respectively 100 and 300 for the GAA and the JAYA algorithm) in order to explore a broad area of the control landscape. For this control problem, we observed that the JAYA algorithm is less efficient than the GAA for the search of a global optimum, but that it is an interesting choice to get quickly (with a small number of iterations of the cost function, and hence, a short computation time) an improved result from a solution which is already quite good.

\bibliographystyle{apsrev4-2}

%

\end{document}